\documentclass[
  aps,
  prl,
  reprint,
  amsmath,
  amssymb,
  footinbib,
  longbibliography,
  superscriptaddress
]{revtex4-2}

\usepackage{graphicx}
\usepackage{dcolumn,amsmath,mathtools}
\usepackage{bm,tikz-cd}

\def\ie{\begin{equation}\begin{aligned}}
\def\fe{\end{aligned}\end{equation}}
\newcommand{\V}{\widehat V}
\newcommand{\U}{\widehat U}
\newcommand{\av}[1]{\left\langle#1\right\rangle}
\newcommand{\Res}{\operatorname*{Res}}
\begin{document}

\title{New relations of tree-level string amplitudes from the massless prescription}

\author{Chen Huang}
\email{chen.huang@soton.ac.uk}
\affiliation{Mathematical Sciences and STAG Research Centre, University of Southampton, Highfield, Southampton, SO17 1BJ, UK}

\author{Yi-Xiao Tao}
\email{tyixiao@itmp.msu.ru}
\affiliation{Department of Mathematical Sciences, Tsinghua University, Beijing 100084, China}
\affiliation{Institute for Theoretical and Mathematical Physics, Lomonosov Moscow State University, Moscow 119991, Russia}
\begin{abstract}
We discover new relations between the string corrections to $(m+1)$-point field-theory amplitudes and $m$-point string amplitudes at tree level from the massless prescription of vertex operators. 
\end{abstract}

\maketitle

\section{Introduction}
The string spectrum includes an infinite tower of massive states. String amplitudes involving these massive states are important for understanding perturbative string theory. Although massive string amplitudes remain relatively unexplored, several previous studies have already produced basic results \cite{Berkovits:2002qx,Bianchi:2015yta,Boels:2012if,Chakrabarti:2017vld,Chakrabarti:2018mqd,Bianchi:2019ywd,Guillen:2021mwp,Huang:2024vrb,Kashyap:2024qor,Huang:2026ekr,Huang:2026uux}.

Beyond the explicit evaluation of string amplitudes, relations among them are important for uncovering deeper structures in string theory. The most famous relation is the Kawai-Lewellen-Tye relation \cite{Kawai:1985xq,Stieberger:2022lss}, which expresses the closed string amplitudes as a special product of two open string amplitudes. The kernel entering this product, known as the KLT kernel, has an elegant mathematical structure \cite{Mizera:2017cqs}. Additional relations have since been discovered, including monodromy relations \cite{Plahte:1970wy, Tourkine:2016bak, Hohenegger:2017kqy} and single-valued maps \cite{Stieberger:2014hba,Schlotterer:2018zce,Brown:2019wna,Gerken:2020xfv,Britto:2021prf,Schlotterer:2025qjv}. Some of these relations can be extended to the AdS string integrals \cite{Alday:2025bjp,Alday:2025cxr,Kakkad:2025klm,Nocchi:2026yms,Tao:2026qeh}. 

In this letter, we will mainly use the pure spinor formalism of string theory \cite{Berkovits:2000fe} and discuss the massless prescription of vertex operators with arbitrary mass level \cite{Chakrabarti:2018mqd}. One of the advantages of the pure spinor formalism is that it manifests the target space supersymmetry so that the string amplitudes are superamplitudes directly, and the pure spinor formalism also allows us to obtain superamplitude relations via a simple procedure. Based on the nested-bracket formalism in the pure spinor superspace \cite{Huang:2026ekr} and the massless prescription, we will show a new string tree amplitude relation involving the $m$-point string tree amplitudes and the $\zeta_k$ coefficients of the low energy expansion of $(m+1)$-point string tree amplitudes which are also known as the $\zeta_k$ multiple-zeta-value (MZV) amplitudes \footnote{The usual definition is based on the $f$-alphabet rather than MZVs. However, for $\zeta_k$ terms here they are equivalent. See \cite{Brown:2011ik, Schlotterer:2012ny, Mafra:2021wok}, }. Since such MZV amplitudes are exactly the string corrections to the corresponding field-theory amplitudes, this relation builds a bridge between $m$-point string amplitudes and the string corrections to $(m+1)$-point field-theory amplitudes. 

In the literature, reference \cite{Kashyap:2023cdi} reported a similar relation between the $\zeta_2$ MZV amplitudes associated with $(m+1)$-point massless string amplitudes and $m$-point string amplitudes with $m-1$ massless vertices and one first-excited state. By comparison, our result applies to $m$-point string amplitudes with arbitrary mass-level distributions and to the $\zeta_k$ MZV amplitudes associated with $(m+1)$-point amplitudes having $m-1$ vertices of arbitrary mass level and two massless vertices, which is the most general case.

\section{Massless prescription of vertices with any mass from OPE}
A mass-level $n$ vertex operator can be described by the OPE between two massless vertices \cite{Friedan:1985ge,Chakrabarti:2018mqd}, and the massless prescription in the pure spinor formalism for unintegrated vertices reads
\ie
V^{(n)}_3=\oint_z\text{d}w U_1^{(0)}(w)V_2^{(0)}(z),\quad 2\alpha'k_1\cdot k_2=-n,
\fe
where $V^{(w)}$ and $U^{(w)}$ are unintegrated and integrated vertices with mass-level $w$ respectively. To manifest the pole structure of the string amplitudes and simplify the expressions, we will use the hatted operator defined below \cite{Huang:2026ekr}:
\ie
V^{(w)}=[\V^{(w)},e^{ik\cdot X}]_0,\, U^{(w)}=[\U^{(w)},e^{ik\cdot X}]_0
\fe
The labels $(w)$ for hatted operators represent the conformal weights: $\V^{(w)}$ has weight $w$ and $\U^{(w)}$ has weight $w+1$. The bracket formalism $[\cdot,\cdot]_i$ is reviewed in the Supplementary Material. The massless prescription of hatted operators can be written as 

\ie
\displaystyle
\widehat V_{3}^{(n)}
=\sum_{\sigma}
&\frac{
\Big[P^{r_1}[\ldots
 \big[P^{r_\ell}
 [\widehat U_1^{(0)}\widehat V_2^{(0)}]_{1-n+q}\big]_0
 \ldots]\Big]_0}
{\ell!\,r_1!\cdots r_\ell!},\\
&\quad\quad\quad\quad\quad\quad\quad\quad\quad\quad2\alpha' k_1\cdot k_2=-n,
\fe
where the sum $\sum_{\sigma}$ stands for $\sum_{q=0}^{n}\sum_{\ell=0}^{q}
\sum_{\substack{r_1,\ldots,r_\ell\ge1\\ r_1+\cdots+r_\ell=q}}$ and $P^{i}\equiv ik_4\!\cdot\!\partial^{i}X$.

As an example, a mass-level 1 vertex can be expressed as
\ie\label{rep1}
\V_3^{(1)}=[\U_1^{(0)}\V_2^{(0)}]_0+[ik_1\cdot\partial X,&[\U_1^{(0)}\V_2^{(0)}]_1]_0, \\
&~~~2\alpha'k_1\cdot k_2=-1,
\fe
which agrees with the results from an explicit calculation in \cite{Kashyap:2024qor}. 

For an arbitrary 3-point amplitude with mass levels of 3 vertices distributed as $(abn)$, if we use the massless prescription to express the vertex with mass level $n$, we expect the result to be the residue of a 4-point amplitude with mass levels $(ab00)$ at $s=-n$, where $s=\alpha'(k_3+k_4)^2$. We will first prove this claim using the hatted operators to show the validity of this formalism, before deriving the string amplitude relations. To simplify the expression, we will omit the label $(w)$, which can be directly read off from the labels of amplitudes.

\section{Expand 3-point string amplitudes with mass levels $(ab1)$}
For the 3-point string amplitude $\mathcal A_{ab1}$, we set $a+b\geq2$ without loss of generality. As mentioned in \cite{Huang:2026ekr}, such an amplitude can be written as \footnote{The extra factor $(-1)^{a+b+1}$ comes from the difference between the usual Koba-Nielsen factor and the one used in \cite{Huang:2026ekr}.}:
\ie\label{3pt1}
\mathcal A_{ab1}(123)=(-1)^{a+b+1}\av{[[\V_1\V_2]_{a+b-1}\V_3]_2}
\fe
Substituting \eqref{rep1} into \eqref{3pt1}, we obtain
\ie
\mathcal A^{\text{massless}}_{ab1}(123)=&(-1)^{a+b+1}
\\
&\times\Big(\av{[[\V_1\V_2]_{a+b-1},[\U_4\V_3]_0]_2}
\\
&-\av{[[\V_1\V_2]_{a+b-1},[\U_4\V_3]_1]_1}
\\
&+(t+a)\av{[[\V_1\V_2]_{a+b}[\U_4\V_3]_1]_0}\Big)
\fe
where $t=\alpha'(k_1+k_4)^2$, $A^{\text{massless}}_{ab1}$ denotes the three-point amplitude $A_{ab1}$ with its third (level-one) leg expanded in massless vertices. After some bracket identities, the final result can be written as:
\ie\label{exp1}
\mathcal A_{ab1}^{\text{massless}}=&(-1)^{a+b+1}
\\
&\times\Big((1-t-a)
\av{[[[\U_4\V_1]_1\V_2]_{a+b},\V_3]_0}
\\
&+\av{[[[\U_4\V_1]_2\V_2]_{a+b-1},\V_3]_0}
\\
&+(-1)^{a+b}(t+a)
\av{[[[\U_4\V_2]_1\V_1]_{a+b},\V_3]_0}
\\
&+(-1)^{a+b}
\av{[[[\U_4\V_2]_2\V_1]_{a+b-1},\V_3]_0}
\\
&+(t+a)\av{[\U_4[[\V_1\V_2]_{a+b}\V_3]_0]_1}
\\
&-\av{[\U_4[[\V_1\V_2]_{a+b-1}\V_3]_1]_1}\Big).
\fe
The last two terms are BRST-exact, as shown in the End Matter, and can be dropped directly since amplitudes are defined on BRST cohomology. 

Now let us turn to the 4-point amplitude with mass levels distributed as $(ab00)$. After some calculations, the final result is
\ie
\mathcal A_{ab00}=&(-1)^{a+b+1}\sum_{n=1}^{a+1}
 B(s+n-1,\,t+a+1-n)
 \\
 &\times\av{[[[\U_4\V_1]_n\V_2]_{a+b+1-n}\V_3]_0}\\
&+\sum_{m=1}^{b+1}(-1)^{m+1}
 B(s+m-1,\,t+a+1)\\
 &
 \times\av{[[[\U_4\V_2]_m\V_1]_{a+b+1-m}\V_3]_0}
\label{eq:integrated-gamma-families}
\fe
with the residue at $s=-1$
\ie
&\text{Res}_{s=-1}\mathcal A_{ab00}\\
=&(-1)^{a+b}(t+a-1)
\av{[[[\U_4\V_1]_1\V_2]_{a+b}\V_3]_0}\\
&-(-1)^{a+b}
\av{[[[\U_4\V_1]_2\V_2]_{a+b-1}\V_3]_0}\\
&-(t+a)
\av{[[[\U_4\V_2]_1\V_1]_{a+b}\V_3]_0}\\
&-\av{[[[\U_4\V_2]_2\V_1]_{a+b-1}\V_3]_0},
\label{eq:recorded-residue}
\fe
which is exactly $\mathcal A_{ab1}^{\text{massless}}$. Here $B(x,y)=\Gamma(x)\Gamma(y)/\Gamma(x+y)$ is the Euler beta function. As for cases $a+b<2$, we have solved them case by case and also found them matching.

\section{Expand 3-point string amplitudes with mass levels $(abn)$}
In this case, we assume that $a+b\geq n+1$ without loss of generality. The 3-point amplitude $(abn)$ reads
\ie
\mathcal A_{abn}(123)=(-1)^{a+b+n}\av{[[\V_1\V_2]_{a+b-n}\V_3]_{2n}}
\fe
and the massless prescription is
\ie
&\mathcal A^{\text{massless}}_{abn}(123)=
\sum_{q=0}^{n}\sum_{k=0}^{n}\sum_{r=0}^{q}
(-1)^q
\binom{t+a}{r}\binom{u+b}{q-r}\binom{-r}{k}
\\
&\times
(-1)^{a+b+n}\av{\left[
 [\V_1\V_2]_{a+b-n+k},
 [\U_4\V_3]_{1-n+q}
 \right]_{2n-q-k}}
\fe
The residue of the 4-point amplitude with mass levels $(ab00)$ at $s=-n$ is
\ie
\Res_{s=-n}\mathcal A_{ab00}
={}&
\sum_{j=1}^{\min(a+1,n+1)}
(-1)^{a+b+n-j}\binom{t+a-j}{n+1-j}
\\
&\times
\av{[[[\U_4\V_1]_j\V_2]_{a+b+1-j},\V_3]_0}\\
&+(-1)^n\sum_{m=1}^{\min(b+1,n+1)}
\binom{t+a}{n+1-m}
\\
&\times
\av{[[[\U_4\V_2]_m\V_1]_{a+b+1-m}\V_3]_0}.
\fe
Applying the bracket identities, we have the following relation:
\ie
\mathcal A_{abn}^{\text{massless}}-&\Res_{s=-n}\mathcal A_{ab00}\\
=&
(-1)^{a+b+n}\sum_{k=0}^{n}(-1)^{n+k}\binom{t+a+k-1}{k}
\\
&\times\av{[\U_4[[\V_1\V_2]_{a+b-n+k}\V_3]_{n-k}]_1}\sim0
\fe
Note that all nested brackets $$[\U_4[[\V_1\V_2]_{a+b-n+k}\V_3]_{n-k}]_1$$ with $0\leq k\leq n$ are BRST-exact and hence can be set to zero. The proof can be found in the end matter. When we take $n=1$, the result matches the $(ab1)$ case discussed above. The case $a+b<n+1$ can also be proved by similar discussions.

Physically, the relation between the massless prescription of $m$-point string amplitudes and the residue of $(m+1)$-point string amplitudes is expected to be valid, i.e.,
\ie
\mathcal A^{\text{massless}}_{w_1w_2\ldots w_{m-1}n}=\Res_{s_{m,m+1}=-n}\mathcal A_{w_1w_2\ldots w_{m-1}00}
\fe
with $s_{m,m+1}=\alpha'(k_m+k_{m+1})^2$. We have verified this by the massless representation of string amplitudes with mass levels $(0002)$ and $(00002)$.

\section{New string amplitude relations from the massless prescription}
The low energy expansion of a string amplitude can be written as
\ie
\mathcal A(P)=A(P)+\sum_{k=2}^{\infty}A^{(\zeta_k)}(P)\zeta_k+\cdots
\fe
where $A(P)$ is the corresponding field-theory amplitude of $\mathcal A(P)$ \footnote{Although $\zeta_{2k}\propto\zeta_2^k$, here we still keep the notation $\zeta_{2k}$ for simplicity.} and we call $A^{(\zeta_k)}(P)$ the $\zeta_k$ MZV amplitude \cite{Brown:2011ik, Schlotterer:2012ny, Mafra:2021wok}. Note that other MZVs are included in the ellipsis. References \cite{Kashyap:2024qor} and \cite{Kashyap:2023cdi} have shown that the residue of a 4-point massless amplitude at $s=-1$ is the same as the $\zeta_2$ MZV amplitudes in its low energy expansion, which is the $F^4$ correction to the SYM amplitudes:
\ie
\Res_{s=-1}\mathcal A_{0000}(1234)&=-\alpha'^2A^{F_4}(1234)\big|_{s=-1}\\
&=A^{\zeta_2}(1234)\big|_{s=-1}.
\fe
This result can be generalized to our $A_{ab00}$ case:
\ie\label{relation1}
\mathcal A_{abn}^{\text{massless}}&=\Res_{s=-n}\mathcal A_{ab00}
=\sum_{k=2}^{n+1}
d_{k,n}\,
A_{ab00}^{(\zeta_k)}\bigg|_{s=-n}.
\fe
The coefficient $d_{k,n}$ can be expressed as
\begin{align}
d_{k,n}
=
\dfrac{s(n+1,k)}{n!}
+\frac{1}{n}\sum_{j=k+1}^{n+1}d_{j,n}\,c_{j,k}(n)
,\,
2\leq k\leq n+1,
\end{align}
with
\ie
&c_m
=
(-1)^{m+1}
\frac{B_{2m}2^{2m-1}6^m}{(2m)!},
\\
&c_{2m,k}(n)
=
\frac{1}{c_m}
\sum_{\ell=1}^{m}
\frac{1}{\ell!}
\sum_{\substack{r_1,\ldots,r_\ell\geq1\\
r_1+\ldots+r_\ell=m}}
\;
\sum_{\substack{s_1,\ldots,s_\ell\geq0\\
s_1+\ldots+s_\ell=k-1}}
\prod_{i=1}^{\ell}
\frac{c_{r_i}}{2r_i}
\\
&\times
\left[
2n^{2r_i}\delta_{s_i,0}
+
(1-\delta_{s_i,0})
\left(
\binom{2r_i}{s_i}(-1)^{s_i}n^{2r_i-s_i}
-\delta_{s_i,2r_i}
\right)
\right],
\\
&c_{2r+1,k}(n)
=
-\frac{1}{2r+1}
\left[
\binom{2r+1}{k-1}(-1)^{k-1}
n^{2r+2-k}
-\delta_{k-1,2r+1}
\right],
\fe
where $B_{2m}$ is the Bernoulli number and the signed Stirling numbers of the first kind $s(n,r)$ is defined by
\begin{equation}
(t-n+1)_n
=\sum_{r=1}^{n}s(n,r)t^r.
\end{equation}
For \(k=n+1\), we have \(d_{n+1,n}=1/n!\). An example of \eqref{relation1} is that, we have $d_{3,2}=1/2$ and $d_{2,2}=-1/2$, and hence
\ie
\mathcal{A}_{ab2}^{\text{massless}}=\frac{1}{2}A_{ab00}^{(\zeta_3)}\big|_{s=-2}-\frac{1}{2}A_{ab00}^{(\zeta_2)}\big|_{s=-2}.
\fe

The relation \eqref{relation1} shows that the massless prescription for a mass level $n$ vertex in 3-point amplitudes leads to a combination of different $\zeta_k$ MZV amplitudes in the 4-point amplitude.

Generally, the residue of an $(m+1)$-point amplitude at $s_{m,m+1}=-n$ cannot be written as a combination of $\zeta_k$ MZV amplitudes, unlike the $n=1$ case \cite{Kashyap:2023cdi}. However, a relation based on multi-factorization of string amplitudes exists. We can factorize the $(m+1)$-point amplitude into a product of some 3-point amplitudes:
\begin{widetext}
\begin{align}
\prod_{j=2}^{m-2}\Res_{s_{1\ldots j}=0}\Res_{s_{m,m+1}=-n}\mathcal A_{w_1w_2\ldots w_{m-1}00}=&\Res_{s_{m,m+1}=-n}\sum_{y_i=\text{massless}}\mathcal A(1,2,y_1)\prod_{j=3}^{m-2}\mathcal A(y_{j-2},j,y_{j-1})\mathcal A(y_{m-3},m-1,m,m+1)\notag\\
=&\sum_{k=2}^{n+1}d_{k,n}\sum_{y_i=\text{massless}}\mathcal A(1,2,y_1)\prod_{j=3}^{m-2}\mathcal A(y_{j-2},j,y_{j-1}) A_{y_{m-3}w_{m-1}00}^{(\zeta_k)}\bigg|_{s_{m,m+1}=-n}\notag\\
=&\sum_{k=2}^{n+1}d_{k,n}\prod_{j=2}^{m-2}\Res_{s_{1\ldots j}=0}A_{w_1w_2\ldots w_{m-1}00}^{(\zeta_k)}\bigg|_{s_{m,m+1}=-n}
\end{align}
\end{widetext}
where $y_i$ runs over all massless polarizations and $s_{12\ldots j}=\alpha'(\sum_{i=1}^jk_i)^2$. The last step follows because 3-point string amplitudes do not provide extra $\zeta_k$-terms. We restrict the cuts to massless intermediate states $y_i$, so that the order of taking residues and performing the low energy expansion can be interchanged freely. We can summarize the following relation between $m$-point and $(m+1)$-point string amplitudes from the massless prescription:
\ie\label{relationm}
\prod_{j=2}^{m-2}&\Res_{s_{1\ldots j}=0}\mathcal A^{\text{massless}}_{w_1w_2\ldots w_{m-1}n}\\
&=\sum_{k=2}^{n+1}d_{k,n}\prod_{j=2}^{m-2}\Res_{s_{1\ldots j}=0}A_{w_1w_2\ldots w_{m-1}00}^{(\zeta_k)}\bigg|_{s_{m,m+1}=-n}.
\fe
Note that such relations also hold for other factorizations. For example, in addition to the residues at $s_{12}=0$ and $s_{123}=0$, we have the following choice for the 6-point string amplitude $A_{w_1w_2w_3w_400}$:
\ie
\Res_{s_{12}=0}\Res_{s_{34}=0}&\,\text{$\Res_{s_{56}=-n}$}\mathcal A_{w_1w_2w_3w_400}\\
&=\Res_{s_{12}=0}\Res_{s_{34}=0}\mathcal A^{\text{massless}}_{w_1w_2w_3w_4n}\\
&=\sum_{k=2}^{n+1}d_{k,n}\Res_{s_{12}=0}\Res_{s_{34}=0}A_{w_1w_2w_3w_400}^{(\zeta_k)}\big|_{s_{56}=-n}.
\fe

\section{Examples}
In this section we give some explicit examples of the relation \eqref{relationm}. We will focus on the residue of the 5-point massless tree-level string amplitude $\mathcal A_{00000}$. This massless 5-point string amplitude can be expressed in terms of SYM amplitudes \cite{Mafra:2011nv}:
\ie
\mathcal A_{00000}(12345)=F_{23}A_{\rm SYM}(12345)+F_{32}A_{\rm SYM}(13245)
\fe
with
\ie
F_{23}
={}&s_{12}s_{34}
\int_0^1 dx\int_0^1dy\,
\frac{x^{s_{45}}y^{s_{12}-1}
(1-x)^{s_{34}-1}(1-y)^{s_{23}}}{(1-xy)^{s_{24}}},\\
F_{32}
={}&s_{13}s_{24}
\int_0^1 dx\int_0^1dy\,
\frac{x^{s_{45}}y^{s_{12}}
(1-x)^{s_{34}}(1-y)^{s_{23}}}{(1-xy)^{s_{24}-1}}.
\fe
Explicit calculations can also give the $\zeta_k$ MZV amplitudes. We list $\zeta_2$ and $\zeta_3$ MZV amplitudes here
\ie
&A^{(\zeta_2)}(12345)=(-s_{34}s_{45}+s_{12}(s_{34}-s_{15}))A_{\rm SYM}(12345)\\
&\quad\quad\quad\quad\quad\quad\quad+s_{13}s_{24}A_{\rm SYM}(13245)\\
&A^{(\zeta_3)}(12345)=\bigg[s_{34}(-s_{12}(s_{12}+2s_{23}+s_{34})+s_{34}s_{45}+s_{45}^2)\\
&\quad\quad\quad\quad\quad\quad\quad+s_{12}^2s_{15}+s_{12}s_{15}^2\bigg]A_{\rm SYM}(12345)\\
&\quad-s_{13}s_{24}(s_{12}+s_{23}+s_{34}+s_{45}+s_{15})A_{\rm SYM}(13245)
\fe
The residue of $\mathcal A_{00000}(12345)$ at $s_{12}=0$, $s_{45}=-n$ is
\ie\label{5residue}
\Res_{s_{12}=0}\Res_{s_{45}=-n}&\mathcal A_{00000}(12345)\\
=(-1)^{n+1}s_{34}&\binom{s_{34}-1}{n-1}\Res_{s_{12}=0}A_{\rm SYM}(12345)\big|_{s_{45}=-n}
\fe
and we list the residues of some MZV amplitudes here
\ie\label{zetaresidue}
\Res_{s_{12}=0}A_{00000}^{(\zeta_2)}(12345)&\big|_{s_{45}=-n}\\
=ns_{34}&\Res_{s_{12}=0}A_{\rm SYM}(12345)\big|_{s_{45}=-n}\\
\Res_{s_{12}=0}A_{00000}^{(\zeta_3)}(12345)&\big|_{s_{45}=-n}\\
=ns_{34}&(n-s_{34})\Res_{s_{12}=0}A_{\rm SYM}(12345)\big|_{s_{45}=-n}\\
\Res_{s_{12}=0}A_{00000}^{(\zeta_4)}(12345)&\big|_{s_{45}=-n}\\
=\frac{n}{4}s_{34}(4s^2_{34}-n&s_{34}+4n^2)\Res_{s_{12}=0}A_{\rm SYM}(12345)\big|_{s_{45}=-n}
\fe
From the definition of $d_{k,n}$, we have
\ie\label{dvalue}
d_{3,2}=\frac{1}{2},d_{2,2}=-\frac{1}{2},d_{4,3}=\frac{1}{6}, d_{3,3}=\frac{3}{8},d_{2,3}=-\frac{55}{24}
\fe
Combining \eqref{5residue}, \eqref{zetaresidue} and \eqref{dvalue}, we can then verify the relation \eqref{relationm} for $\mathcal A_{000n}^{\text{massless}}$ at $n=2,3$:
\ie
\Res_{s_{12}=0}&\mathcal A_{0002}^{\text{massless}}=s_{34}(1-s_{34})\Res_{s_{12}=0}A_{\rm SYM}(12345)\big|_{s_{45}=-2}\\
&=\Res_{s_{12}=0}\big(\frac{1}{2}A^{(\zeta_3)}_{00000}-\frac{1}{2}A^{(\zeta_2)}_{00000}\big)\big|_{s_{45}=-2}\\
\Res_{s_{12}=0}&\mathcal A_{0003}^{\text{massless}}\\
&=\frac{s_{34}(s_{34}-1)(s_{34}-2)}{2}\Res_{s_{12}=0}A_{\rm SYM}(12345)\big|_{s_{45}=-3}\\
&=\Res_{s_{12}=0}\big(\frac{1}{6}A^{(\zeta_4)}_{00000}+\frac{3}{8}A^{(\zeta_3)}_{00000}-\frac{55}{24}A^{(\zeta_2)}_{00000}\big)\big|_{s_{45}=-3}\,.
\fe

\section{Conclusion and outlook}
In this letter, we found a relation between $m$-point string amplitude and $(m+1)$-point $\zeta_k$ MZV amplitudes using the massless prescription and proper factorizations. This relation shows that the information of the string correction to $(m+1)$-point field-theory amplitudes is encoded in $m$-point string amplitudes. The massless prescription follows from the worldsheet OPE, suggesting that
this amplitude relation is already encoded in the underlying CFT.

It would be interesting to apply the massless prescription to more vertices in string amplitudes. Since the relation we demonstrated in the letter involves only MZV amplitudes with $\zeta_k$, multiple substitutions may yield analogous relations for higher-depth MZV amplitudes associated with products $\zeta_{k_1}\zeta_{k_2}\cdots\zeta_{k_m}$. Furthermore, similar relations may hold for all MZV amplitudes. Meanwhile, if such relations exist, we can relate any string amplitude to massless MZV amplitudes.

\begin{acknowledgments}
\noindent{\textit{Acknowledgment.---}}DeepSeek-Flash and GPT-6 Astra were used to assist with the analysis, and the authors are responsible for all results. We thank Carlos Mafra for the useful discussion. YT is supported by the National Natural Science Foundation of China (NSFC) (Grant No.~124B2094).

\end{acknowledgments}

\onecolumngrid
\begin{center}
    \textbf{End Matter}
\end{center}
\twocolumngrid
\noindent\textbf{\textit{Bracket identities.---}}
The OPE of operators $A$ and $B$ is defined as:
\begin{equation}
 A(z)B(w)=\sum_{n\leq N}\frac{[AB]_n(w)}{(z-w)^n}.
\end{equation}
Thus $[AB]_1$ is the residue, $[AB]_0$ is the normal-ordered
product, and $[AB]_{-r}=\frac{1}{r!}[\partial^r AB]_0$ for $r\geq0$.
Writing $|A|$ for the Grassmann parity of $A$, the two identities
used to rearrange nested brackets are \cite{Kashyap:2024qor}:
\begin{widetext}
\ie
 \relax[A[BC]_n]_m
 ={}&(-1)^{|A||B|}[B[AC]_m]_n+\sum_{j=0}^{m-1}\binom{m-1}{j}
       [[AB]_{m-j}C]_{n+j}, \, m\geq1
       \\
 [[AB]_mC]_n
 ={}&\sum_{j\geq0}(-1)^j\binom{m-1}{j}
 \Bigl([A[BC]_{n+j}]_{m-j}+(-1)^{m+|A||B|}
       [B[AC]_{j+1}]_{m+n-j-1}\Bigr).
\fe
\end{widetext}
The second identity includes $m=0$ with $\binom{-1}{j}=(-1)^j$.
Exchanging the two entries of a bracket gives:

\ie
 \relax[BA]_n&=(-1)^{n+|A||B|}
 \sum_{j\geq0}\frac{(-1)^j}{j!}\,
 \partial^j[AB]_{n+j}
 \\
[\partial AB]_n &= (1-n)[AB]_{n-1}
\\
[A\partial B]_n &= \partial[AB]_n + (n-1)[AB]_{n-1} &
\\
[AB]_{n-i}&=\frac{(-1)^i}{i!\binom{n-1}{i}}[\partial^i A B]_n &
\fe
where $n\geq0$. All sums truncate at the highest OPE pole.

\noindent\textbf{\textit{BRST-exactness of $[\U_4[[\V_1\V_2]_{a+b-n+k}\V_3]_{n-k}]_1$.---}}
We use the method in \cite{Huang:2026ekr} to show that all nested-brackets with the form $[\U_4[[\V_1\V_2]_{a+b-n+k}\V_3]_{n-k}]_1$ are BRST-exact here, with the mass levels $(ab00)$. Since we have the hatted to unhatted dictionary
\ie
QU=\partial V\to \partial \hat{V}+[ik\cdot\partial X, \hat{V}]_0
\fe
where $Q$ is the BRST charge, we have
\ie
&0\sim Q[U_4[[V_1U_2]_{a+b-n+k}V_3]_{n-k+1}]_1\\
&\to (n-k)[\U_4[[\V_1\V_2]_{a+b-n+k}\V_3]_{n-k}]_1\\
&-(a+b-n+k-1)[\U_4[[\V_1\V_2]_{a+b-n+k-1}\V_3]_{n-k+1}]_1\\
&-(t+b)[\U_4[[\V_1 \V_2]_{a+b-n+k}\V_3]_{n-k}]_1\\
&+(s+a+b)[\U_4[[\V_1 \V_2]_{a+b-n+k-1}\V_3]_{n-k+1}]_1\sim 0
\fe
for $n-k\geq -1$. Then we get a recursion relation
\ie
&(s+n-k+1)[\U_4[[\V_1 \V_2]_{a+b-n+k-1}\V_3]_{n-k+1}]_1\\
&-(t+b-n+k)[\U_4[[\V_1 \V_2]_{a+b-n+k}\V_3]_{n-k}]_1\sim 0
\fe
If we take $n-k=-1$, then from $[\V_1 \V_2]_{a+b+1}=0$, we have
\ie
s[\U_4[[\V_1 \V_2]_{a+b}\V_3]_{0}]_1\sim 0
\fe
Hence for general $s$ and $t$, all nested brackets $[\U_4,[[\V_1\V_2]_{a+b-n+k},\V_3]_{n-k}]_1$ are BRST-exact and can be set to zero. Since there is no Koba-Nielsen factor included in the nested brackets constructed from hatted operators, such brackets can be regarded as continuous functions of $s$ and $t$. By continuity, they can also be set to zero even when $s$ and $t$ take special values such that $s+n-k+1$ or $t+b-n+k$ vanishes. This fact has been verified by an explicit calculation with mass levels $(1100)$.

\bibliographystyle{apsrev4-1}
\bibliography{abn}

\end{document}